\documentclass[12pt]{article}
\usepackage{epsf,latexsym,rotate}
\usepackage[ps,arc,frame]{xy}
\usepackage{amsfonts,amssymb} 
\epsfverbosetrue
\usepackage[ps,arc,frame]{xy}
\usepackage{graphicx}
\usepackage{slashed}

\newcommand{\lapprox}{%
\mathrel{%
\setbox0=\hbox{$<$}
\raise0.6ex\copy0\kern-\wd0
\lower0.65ex\hbox{$\sim$}
}}
\newcommand{\gapprox}{%
\mathrel{%
\setbox0=\hbox{$>$}
\raise0.6ex\copy0\kern-\wd0
\lower0.65ex\hbox{$\sim$}
}}

\newcommand{\goth}[1]{\mathfrak{#1}}
\newcommand{\double}[1]{\mathbb{#1}}

\newcommand{\cc}{\double{C}}
\newcommand{\nn}{\double{N}}
\newcommand{\rr}{\double{R}}
\newcommand{\zz}{\double{Z}}

\newcommand{\aaa}{\mathcal{A}}
\newcommand{\ccc}{\mathcal{C}}

\newcommand{\gggg}{\goth{g}}
\newcommand{\hhh}{\double{H}}
\newcommand{\mm}{\mathcal{M}}

\newcommand{\pp}{\pmatrix}

\newcommand{\dd}{\mathcal{D}}

\newcommand{\hh}{\mathcal{H}}

\newcommand{\de}{\hbox{\rm{d}}}

\newcommand{\pa}{\partial}

\newcommand{\lb}{\left[}
\newcommand{\rb}{\right]}

\newcommand{\ul}[1]{\underline{#1}}
\newcommand{\ot}{\otimes}
\newcommand{\op}{\oplus}

\newcommand{\bb}{\begin{eqnarray}}
\newcommand{\ee}{\end{eqnarray}}
\newcommand{\eee}{\nonumber\end{eqnarray}}
\newcommand{\qq}{\quad}

\newcommand{\minibox}[2]{{\fbox{\begin{minipage}{#1}\begin{center}#2
\end{center}\end{minipage}}}}

\newcommand{\rxy}[1]{{\begin{xy}
0;<2mm,0mm>:<0mm,2mm>::0;0,#1\end{xy}}}

\newcommand{\rxyh}[2]{{\begin{xy} 0;<2mm,0mm>:<0mm,2mm>::0;0,
,(5,-2)*{a}
,(10,-2)*{b}
,(15,-1.8)*{\bar{b}}
,(20,-2)*{c}
,(25,-1.8)*{d}
,(30,-1.8)*{\bar{d}}
,(2,-5)*{a}
,(2,-10)*{b}
,(1.8,-15)*{\bar{b}}
,(2,-20)*{c}
,(1.8,-25)*{d}
,(1.8,-30)*{\bar{d}}
,(5,-5)*\cir(#1,0){}
,(10,-5)*\cir(#1,0){}
,(15,-5)*\cir(#1,0){}
,(20,-5)*\cir(#1,0){}
,(25,-5)*\cir(#1,0){}
,(30,-5)*\cir(#1,0){}
,(5,-10)*\cir(#1,0){}
,(10,-10)*\cir(#1,0){}
,(15,-10)*\cir(#1,0){}
,(20,-10)*\cir(#1,0){}
,(25,-10)*\cir(#1,0){}
,(30,-10)*\cir(#1,0){}
,(5,-15)*\cir(#1,0){}
,(10,-15)*\cir(#1,0){}
,(15,-15)*\cir(#1,0){}
,(20,-15)*\cir(#1,0){}
,(25,-15)*\cir(#1,0){}
,(30,-15)*\cir(#1,0){}
,(5,-20)*\cir(#1,0){}
,(10,-20)*\cir(#1,0){}
,(15,-20)*\cir(#1,0){}
,(20,-20)*\cir(#1,0){}
,(25,-20)*\cir(#1,0){}
,(30,-20)*\cir(#1,0){}
,(5,-25)*\cir(#1,0){}
,(10,-25)*\cir(#1,0){}
,(15,-25)*\cir(#1,0){}
,(20,-25)*\cir(#1,0){}
,(25,-25)*\cir(#1,0){}
,(30,-25)*\cir(#1,0){}
,(5,-30)*\cir(#1,0){}
,(10,-30)*\cir(#1,0){}
,(15,-30)*\cir(#1,0){}
,(20,-30)*\cir(#1,0){}
,(25,-30)*\cir(#1,0){}
,(30,-30)*\cir(#1,0){}
#2\end{xy}}}

\begin{document}

\font\twelve=cmbx10 at 13pt
\font\eightrm=cmr8

\thispagestyle{empty}

\begin{center}

CENTRE DE PHYSIQUE TH\'EORIQUE $^1$ \\ CNRS--Luminy, Case
907\\ 13288 Marseille Cedex 9\\ FRANCE\\

\vspace{2cm}

{\Large\textbf{On the noncommutative standard model}} \\

\vspace{1.5cm}

{\large Jan-H. Jureit $^2$, Thomas Krajewski $^3$,\\ Thomas
Sch\"ucker $^4$, Christoph A. Stephan $^5$}

\vspace{2cm}

{\large\textbf{Abstract}}
\end{center}
We propose a pedestrian review of the noncommutative standard model
in its present state. \vspace{1.2cm}

${}$\hfil {\it dedicated to Alain Connes on the occasion of his 60th  birthday}\break
\vskip 1,0 truecm

\noindent
PACS-92: 11.15 Gauge field theories\\
MSC-91: 81T13 Yang-Mills and other gauge theories

\vskip 1truecm


\vspace{1.5cm}
\noindent $^1$ Unit\'e Mixte de Recherche  (UMR 6207)
du CNRS  et des Universit\'es Aix--Marseille 1 et 2 et  Sud
Toulon--Var, Laboratoire affili\'e \`a la FRUMAM (FR 2291)\\
$^2$  also at Universit\"at Kiel, jureit@cpt.univ-mrs.fr\\
$^3$  also at Universit\'e Aix--Marseille 1,
krajew@cpt.univ-mrs.fr\\
$^4$ also at Universit\'e Aix--Marseille 1,
thomas.schucker@gmail.com \\
$^5$ also at Universit\'e Aix--Marseille 1,
christophstephan@gmx.de\\

\section{Introduction}

Understanding the origin of the standard model is currently one of
most challenging issues in high energy physics. Indeed, despite its
experimental successes, it is fair to say that its structure remains
a mystery. Moreover, a better understanding of its structure would
provide us with a precious clue towards its possible extensions.

This can be achieved in the framework of noncommutative geometry
\cite{book}, which is a branch of mathematics pioneered by Alain Connes
and  aiming at a generalization of geometrical ideas to spaces whose
coordinates fail to commute. Motivated by quantum gravity, it is
postulated that space-time is a wildly noncommutative manifold at a
very high energy. Even if the precise nature of this noncommutative
manifold remains unknown, it seems legitimate to assume that at an
intermediate scale, say a few orders of magnitude below the Planck
scale, the corresponding algebra of coordinates is only a mildly
noncommutative algebra of matrix valued functions. When suitably
chosen, such a matrix algebra reproduces within the spectral action
principle the standard model coupled to gravity \cite{cc}.

It is worthwhile to notice that this is a {\it bottom-up} approach,
as opposed to string theory which is a {\it top-down} one. Indeed,
in noncommutative geometry, one tries to guess the small scale
structure of space-time from our present knowledge at the
electroweak scale, whereas string theory aims at deriving the
standard model directly from the Planck scale physics.

Nevertheless, the physical interpretation of the spectral action
principle and its confrontation with present-day experiment still
require some contact with the low energy physics. This follows from
the standard Wilsonian renormalization group idea. The spectral
action provides us with a bare action supposed to be valid at a
very high energy of the order of the unification scale. Then,
evolving down to the electroweak scale yields the effective low
energy physics. This line of thought is very similar to the one
adopted in grand unified theories. Indeed, in a certain sense
models based on non commutative geometry can be considered as
alternatives to grand unification that do not imply proton decay.

Ten years after its discovery \cite{cc}, the spectral action has  recently received new impetus \cite{c06,barr,mc2} by allowing a  Lorentzian signature in the internal space. This mild modification  has three consequences.
\begin{itemize}\item
The fermion-doubling problem \cite{2f} is solved elegantly.
\item
Majorana masses and consequently the popular seesaw mechanism are  allowed for.
\item
The Majorana masses in turn decouple the Planck mass from the $W$ mass.
\end{itemize}
Furthermore, Chamseddine, Connes \& Marcolli point out an additional  constraint on the coupling constants tying the sum of all Yukawa  couplings squared to the weak gauge coupling squared. This relation  already holds for Euclidean internal spaces
\cite{thum}.

The aim of this paper is to review the present status of the  noncommutative standard model in a pedestrian fashion and to  illustrate it by numerical examples.

\subsection{A Flavour of Noncommutative Geometry}

The unification of gravity with the forces of the sub-atomic world,  i.e. the electroweak and
the strong force is one of the major problems of theoretical physics.
The non-gravitational
forces are coded in the standard model of particle physics. Taking  quantum
mechanics as a basis, the standard model is formulated as a quantum  field theory, which
perturbatively gives extraordinarily precise experimental predictions.  From the differential geometric point of
view, the total gauge group $G=SU(2)\times U(1)\times SU(3)/(\zz_2 \times \zz_3)$ acts on the matter fields
being sections in a vector bundle associated to the principal $G$-bundle over the space-time
manifold. Since gauge transformations are taken
locally, one can thus imagine an 'internal' space which is attached  to each
point of the manifold, i.e. reflecting the gauge degrees of freedom.  This 'gauge space' is similar to Kaluza's point of view.

The geometry of general relativity is different,
Riemannian geometry of a
curved space-time. By the equivalence principle, gravity arises as a  pseudo-force
from a general coordinate transformation. General coordinate transformations
are diffeomorphisms of the underlying Riemannian manifold, leaving as  such
the Einstein-Hilbert action invariant.  The symmetry acts directly on  the manifold $M$ and on its metric.

Would it not be nice to have a picture for the standard model, where
symmetries act 'directly' on an underlying space-time manifold
producing the electroweak and strong forces as pseudo-forces by a  group of coordinate transformations?
Would it not be nice to place
gravity and the other forces on the same footing, namely by
obtaining {\it all} forces as pseudo-forces from some general
'coordinate transformations' acting on some general 'space-time'?
This is what Alain Connes' noncommutative
geometry \cite{book} does for you.

In quantum mechanics, points of the phase space lose their meaning
due to Heisenberg's uncertainty relation, $\Delta x\Delta p\ge \hbar/2$.
The commutative algebra of classical observables, i.e. functions on  phase
space, is made into a noncommutative involution algebra $\mathcal{A}$  due to
$[\hat{x}_i,\hat{p}_j]= i \hbar\delta_{ij}$, with the involution $ \cdot^\ast$ being
the Hermitean conjugation. In the relativistic setting, the  wavefunctions
are square integrable spinors living in the Hilbert space
$\mathcal{H}=\mathcal{L}^2(\mathcal{S})$. The algebra $\mathcal{A}$  is faithfully
represented on $\mathcal{H}$ and the dynamics is given by the Dirac
operator $\slashed\partial\in\mathrm{End}(\mathcal{H})$.
It is this `spectral' triple $(\mathcal{A},\mathcal{H},\slashed \partial)$ (with some
additional structure) which describes  a Connes' geometry. A slight  shortcoming
is the requirement for an Euclidean setting in order to have $\slashed \partial^\ast=\slashed\partial$. One assumes that this can be
cured by a Wick rotation, but it is still an open question how to
implement a Lorentzian signature.
\begin{figure}[h]
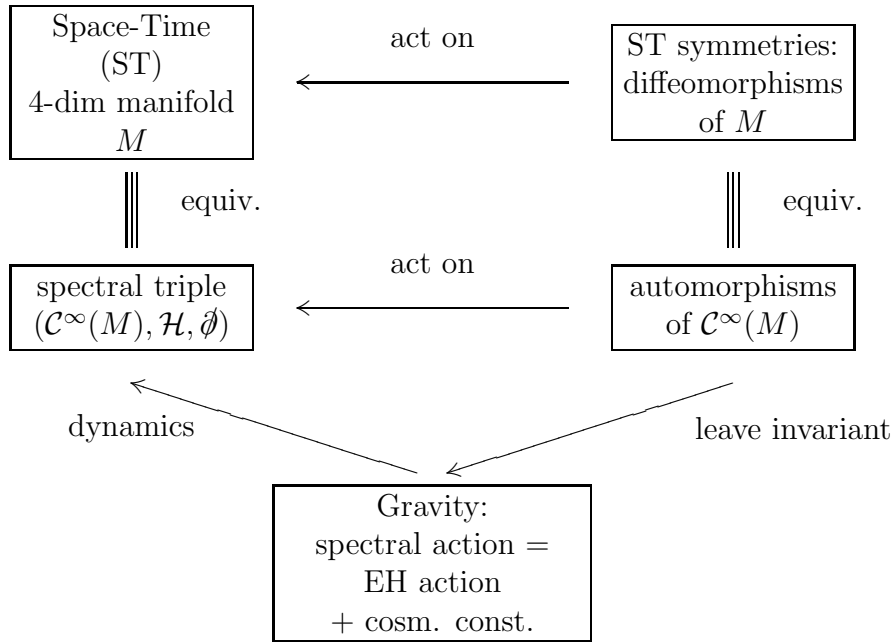

\centering
\begin{tabular}{c}
\\
\rxy{
,(-20,15)*{
\minibox{3cm}{Space-Time (ST)\\ 4-dim manifold $M$}
}
,(20,15)*{
\minibox{3cm}{ST symmetries: \\ diffeomorphisms of $M$}
}
,(-9,15);(9,15)**\dir{-}?(0)*\dir2{<}
,(-20,9);(-20,4)**\dir3{-}
,(20,9);(20,4)**\dir3{-}
,(-14,7)*{\mbox{equiv.}}
,(26,7)*{\mbox{equiv.}}
,(0,18)*{\mbox{act on}}
,(-20,0)*{
\minibox{3cm}{spectral triple\\ $(\ccc^\infty(M),\mathcal{H},\slashed {\partial}
)$}
}
,(20,0)*{
\minibox{3cm}{ automorphisms \\ of $\ccc^\infty(M)$}
}
,(0,-17)*{
\minibox{4cm}{Gravity: \\ spectral action = \\ EH action\\+ cosm.  const.}
}
,(-20,-5);(-1,-11)**\dir{-}?(0)*\dir2{<}
,(20,-5);(1,-11)**\dir{-}?(1)*\dir2{>}
,(-9,0);(9,0)**\dir{-}?(0)*\dir2{<}
,(0,3)*{\mbox{act on}}
,(24,-8)*{\mbox{leave invariant}}
,(-20,-8)*{\mbox{dynamics}}
}
\\ \\
\end{tabular}
\caption{Gravity from commutative geometries}
\label{spec1}
\end{figure}

Of particular interest are the commutative spectral triples coming
from a compact spin manifold $M$ with
$\aaa=\ccc^\infty(M)$, $ \mathcal{H}=\mathcal{L}^2(\mathcal{S})$  and  $\slashed\partial$. The axioms of the spectral triple are such that  there is a one-to-one correspondence between commutative, real  spectral triples and Riemannian spin geometries \cite{av}. This  reconstruction theorem tells us in particular how to reconstruct the  Riemannian metric from the operator algebraic data of the spectral  triple.
The diffeomorphisms of the manifold have their equivalent in
the automorphisms of $\mathcal{A}$ lifted to the spinors and
the so called spectral action due to Chamsedinne \& Connes \cite {grav,cc}
reproduces from the eigenvalues of the Dirac operator the Einstein-Hilbert
action with positive cosmological constant.
One can see the general setting in figure \ref{spec1}.

It is now possible to relax the commutativity of the algebra $\mathcal {A}$. In that
sense, a spectral triple $(\mathcal{A},\mathcal{H},\mathcal{D})$ with  a noncommutative algebra
will still be equivalent to some 'manifold', now promoted to a space,  where
points lose their meaning, i.e. a noncommutative space. Connes' geometry
thus does to space-time what
quantum mechanics does to phase space. In particular, it is even  applicable to
discrete spaces, or spaces which have dimension zero. Spectral  triples are
thus versatile enough to describe spaces, noncommutative or not,  discrete or continuous,
on an equal footing. For example, to find an algebra $\mathcal{A}_f$  whose automorphisms
reproduce the gauge symmetries of the standard model, one can define  a {\it finite}
spectral triple, for example with an algebra $\mathcal{A}_f$ being a  direct
sum of matrix algebras. In the case of the standard model, this is
$\mathcal{A}_f=\mathbb{H}\oplus\mathbb{C}\oplus M_3(\mathbb{C})$.

An almost commutative geometry is defined to be the tensor product of  two
spectral triples, the first one describing a 4-dimensional space-time,
and the second is a 0-dimensional discrete spectral triple.
A 0-dimensional triple has a finite dimensional algebra
$\mathcal{A}_f$ and
a finite
dimensional Hilbert space $\mathcal{H}_f$.
\begin{figure}[h]
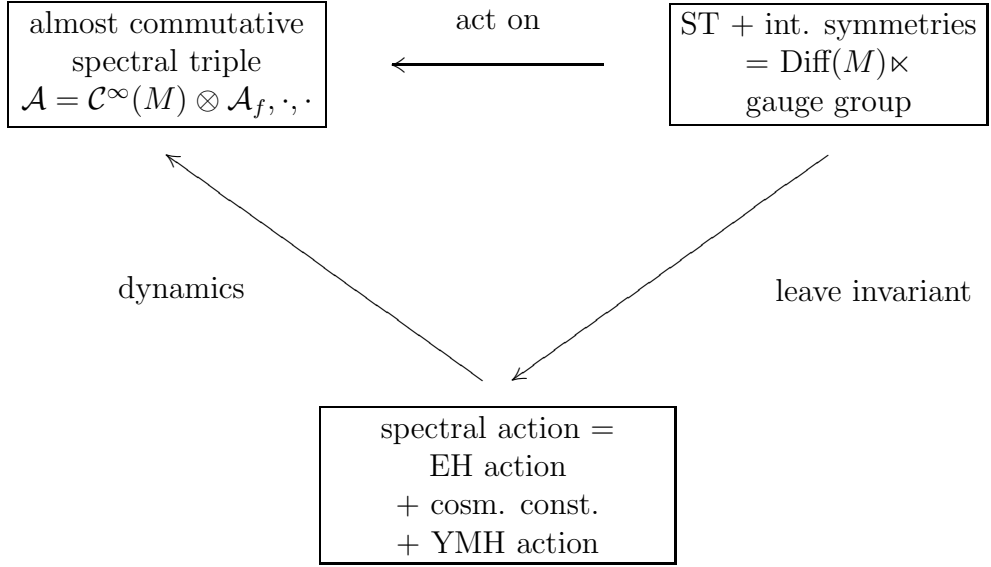

\centering
\begin{tabular}{c}
\rxy{
,(-22,0)*{
\minibox{4cm}{almost commutative \\ spectral triple \\ $\mathcal{A}=  \ccc^\infty(M) \otimes
\mathcal{A}_f,\cdot,\cdot$}
}
,(22,0)*{
\minibox{4cm}{ST + int. symmetries \\ = Diff$(M) \ltimes$ \\ gauge  group}
}
,(0,-28)*{
\minibox{4.5cm}{spectral action = \\ EH action\\ + cosm. const. \\ +  YMH action }
}
,(-22,-6);(-1,-21)**\dir{-}?(0)*\dir2{<}
,(22,-6);(1,-21)**\dir{-}?(1)*\dir2{>}
,(-7,0);(7,0)**\dir{-}?(0)*\dir2{<}
,(0,3)*{\mbox{act on}}
,(25,-15)*{\mbox{leave invariant}}
,(-21,-15)*{\mbox{dynamics}}
}
\\ \\
\end{tabular}
\caption{Gauge forces from almost commutative geometries}
\label{acsm}
\end{figure}
One can show that the spectral action gives the Einstein-Hilbert  action together
with the bosonic part of the standard model action, in the case where  the finite
algebra $\mathcal{A}_f$ is taken to be a direct sum of matrix algebras,
$\mathcal{A}_f=\mathbb{H}\oplus\mathbb{C}\oplus M_3(\mathbb{C})$.
An almost commutative geometry can be viewed as an ordinary  (commutative)
4-dimensional space-time with an 'internal' Kaluza-Klein space  attached to
each point. The 'fifth' dimension is a discrete, 0-dimensional space.
In figure \ref{acsm}, we tried to visualise this geometric landscape.  The automorphisms
are the semi-direct product of Diff$(M)$ and the gauge  transformations acting on the finite
particle content in a certain representation.
It is a rather amazing fact that, since everything is formulated in a  pure
geometrical language, the Higgs fields turns out to be a connection  on the
'internal' space and comes out automatically because of the interplay
between the two algebras $\ccc^\infty(M)$ and $\aaa_f$. Probably, this
is one of the most appealing features of almost commutative geometry.
Indeed, in this framework, the Higgs can be understood as the
'internal' metric with its dynamics given by the Higgs potential.
Calculating the spectral action produces the complete Yang-Mills- Higgs action
of the standard model coupled to gravity.

\section{The problems of the standard model}

It is fair to compare the successes and the shortcomings of the  standard model with those of the Balmer-Rydberg formula:
\bb\nu =g_1n_1^{q_1}+g_2n_2^{q_2}.\label{ansatz}\ee
This ansatz contains two discrete parameters $q_1,q_2\in \zz$ and two  continuous parameters $g_1,g_2\in \rr$, while the $n_i\in\nn$ are  simply labels. These parameters are determined successfully by  fitting the ansatz to atomic spectra. As this phenomenological  success increased so did the urgency to answer three questions:
\begin{itemize}\item
Who ordered the ansatz?
\item
Does this order imply constraints on the discrete parameter?
\item
Does this order imply constraints on the continuous parameter?
\end{itemize}
The three answers came later: quantum mechanics did; yes, $q_1=q_2=-2 $; yes,
\bb g_1=-g_2=\,\frac{m_e}{4\pi \hbar^3}\,\frac{e^4}{(4\pi \epsilon
_0)^2}\ee
for the hydrogen atom in agreement with the experimental fit.

Twisting the history of general relativity only slightly, we can  start with the purely field theoretic ansatz:
\bb S[g]=\int_M\,{\textstyle\frac{-1}{16\pi G}}\, [2\Lambda _c+ R^q
]\,\de V \label{eh},\ee
with one discrete and two continuous parameters, $G$ and $\Lambda_c$.  The three answers are: Riemannian geometry did; not really, but $q=1$  certainly is the cheapest order; no. Those of you who hate geometry  might want to replace the first answer by: Since the meter was  officially abolished on October 21st 1983 we must not use any ansatz relying on  inertial coordinates and the Einstein-Hilbert ansatz (\ref{eh}) with  $q=1$ is the first not to rely on that background.

Now to the non-gravitational forces.  The ansatz is an action
proposed in bits by Klein, Gordon, Dirac, Weyl, Majorana, Yukawa,
Yang, Mills, Brout, Englert and Higgs. Its discrete parameters are a
compact, real Lie group $G$, the `gauge group', and three unitary
representations on complex Hilbert spaces, $\hh_S$ for the Higgs
scalar, $\hh_L$ for the left- and $\hh_R$ for the right-handed Weyl
spinors. The continuous parameters are the gauge-invariant couplings
whose number increases with the number of simple components in the
Lie algebra $\gggg$ of $G$ and increases sharply with the number of
irreducible components in the representations.

The experimental fit yields the following discrete parameters:
\bb G&=&SU(2)\times U(1)\times
SU(3)/(\zz_2\times\zz_3),\label{smgr}\\ \cr
\hh_L &=& \bigoplus_1^3\lb
(2,{\textstyle\frac{1}{6}},3)\op
(2,-{\textstyle\frac{1}{2}},1)
\rb  ,\label{hl}\\
\hh_R& = &\bigoplus_1^3\lb
(1,{\textstyle\frac{2}{3}},3)\oplus
(1,-{\textstyle\frac{1}{3}},3)\op (1,-1,1)\op (1,0,1)
\rb,\label{hr} \\
\hh_S &= &(2,-{\textstyle\frac{1}{2}},1)\label{hs}.
\ee

Consequently we have the following continuous parameters: three gauge  couplings, one quadratic and one quartic Higgs self-coupling, which  are traded for the $W$ and the Higgs masses,
and a bunch of complex Yukawa couplings, which are traded for the 12  Dirac masses, three Majorana masses and three unitary mixing  matrices. We therefore have $3+2+12+3+3 \times 4=32$ physically  relevant, real, continuous parameters.
$4\cdot 3\times 3$ complex Yukawa couplings, which are traded for the  12 Dirac masses and two unitary mixing matrices. Each of these two  contains three angles and one phase. Then there is the complex,  symmetric $3\times 3$ matrix of Majorana masses containing $2\cdot 6$  real parameters, all physical except for one.
We therefore have $3+2+12+2 \times 4 +11=36$ physically relevant,  real, continuous parameters.
Today, many of them are known experimentally with good precision \cite {data}. We do not know much about the last eleven, the Majorana mass  matrix is constrained weakly by neutrinoless double $\beta$-decay.

Now we face the three questions: where does the complicated ansatz  come from, where do its discrete parameters come from, where do its  continuous parameters come from?

\section{Three answers}

\subsection{ Who ordered the ansatz?}

Noncommutative geometry did.
Chamsedine \& Connes have computed the spectral action for a generic  almost commutative geometry. They obtain in addition to the Einstein- Hilbert Lagrangian the following terms:
\begin{itemize}\item
the Yang-Mills Lagrangian,
\item
the Klein-Gordon Lagrangian with the covariant derivative,
\item
the Higgs potential with its spontaneous symmetry breaking,
\item
the Dirac Lagrangian for the left- and right-handed Weyl spinors with  the covariant derivative,
\item
the Yukawa couplings.
\end{itemize}
Note in particular that the first three Lagrangians come with the  correct signs relative to the Einstein-Hilbert Lagrangian. In  addition there is a term quadratic in the Weyl curvature and  a  coupling between curvature scalar and the Higgs scalar making the  scale-invariant part of the Lagrangian invariant under local  dilatations.

We conclude that almost commutative geometry unifies gravity with the  other forces in the sense that the latter become pseudo-forces  accompanying the former. This is similar to Minkowskian geometry  allowing us to interpret the magnetic force as a pseudo-force  accompanying the electric force. Or Riemannian geometry allowing, via  the equivalence principle, to interpret gravity as a pseudo-force.  The definition of a pseudo-force comes with a transformation, a  Lorentz transformation for Minkowskian, a general coordinate  transformation for Riemannian geometry. The corresponding  transformation for noncommutative geometry will be discussed next.

\subsection{Constraints on the discrete parameters}

There are two ways to extract the gauge group $G$ from the spectral  triple.

The first way defines the gauge group to be the unimodular (i.e. of  unit determinant) unitary group \cite{real,mc2} of the associative  algebra, which by the faithful representation immediately acts on the  Hilbert space.

The second way follows general relativity whose invariance group is  the group of diffeomorphisms of $M$ (general coordinate  transformations). In the algebraic formulation this is the group of  algebra automorphisms. Indeed Aut$(\ccc^\infty(M))={\rm Diff}(M).$ We  still have to lift the diffeomorphisms to the Hilbert space. This  lift is double-valued and its image is the semi-direct product of the  diffeomorphism group with the local spin group \cite{lift}. Also it  can be shown perturbatively that in the commutative case the spin  lift is unique \cite{unique}. In some almost commutative cases, the  lift has to be centrally extended in order to remain finitely valued  \cite{fare}. These extensions are not unique but parameterized by  central charges. Let us note that all automorhisms (in the connected  component of the identity) of the associative algebras of inner  spectral triples are inner while the ones of commutative spectral  triples are outer.

Up to possible central $U(1)$s and their central charges, the two  approaches coincide.

It follows that all four infinite series in Cartan's classification
of simple Lie algebras are induced from finite spectral triples, but
not the five exceptional ones. For example, $G_2$ is the Lie algebra
of the automorphism group of the {\it non-}associative algebra of
octonions. This restriction to the gauge groups $U(N)$, $SO(N)$ and
$SP(N)$ is reminiscent from open string theories with gauge fields
arising from the Chan-Paton factors.

In the even-dimensional case the Hilbert space of the spectral triple  is decomposed by the chirality operator into a left- and a right- handed piece, $\hh=\hh_L\op\hh_R$. They define immediately the  unitary representation of the ansatz. However not every group  representation extends to an algebra representation, only the  fundamental ones do. For example take $\aaa=\hhh$, the quaternions.  Its automorphism group is Aut$(\hhh)=SU(2)/\zz_2$, its unitary group  $U(\aaa):=\{u\in\aaa,u^*u=uu^*=1\}=SU(\hhh)=SU(2)$ is already  unimodular. There is only one irreducible representation of the  quaternions, $\hh=\cc^2$. All group representations of $SU(2)$ with  higher spin do not extend to an algebra representation.

There are other constraints on the fermionic representations coming
 from the axioms of the spectral triple. They are conveniently
captured in Krajewski diagrams which classify all possible finite  dimensional spectral triples \cite{kps}. They do for spectral triples  what the Dynkin and weight diagrams do for groups and  representations. Figure 3 shows the Krajewski diagram of the standard  model in Lorentzian signature with one generation of fermions  including a right-handed neutrino.

\begin{center}
\begin{tabular}{c}
\rxyh{0.7}{
,(5,-20)*\cir(0.3,0){}*\frm{*}
,(5,-25)*\cir(0.3,0){}*\frm{*}
,(5,-20);(10,-20)**\dir{-}?(.4)*\dir{<}
,(5,-20);(15,-20)**\crv{(10,-17)}?(.4)*\dir{<}
,(5,-25);(15,-25)**\crv{(10,-28)}?(.4)*\dir{<}
,(5,-25);(30,-25)**\crv{~*=<2pt>{.}(17.5,-20)}?(.45)*\dir{<}
,(20,-5)*\cir(0.3,0){}*\frm{*}
,(25,-5)*\cir(0.3,0){}*\frm{*}
,(20,-5);(20,-10)**\dir{-}?(.6)*\dir{>}
,(20,-5);(20,-15)**\crv{(17,-10)}?(.6)*\dir{>}
,(25,-5);(25,-15)**\crv{(28,-10)}?(.6)*\dir{>}
,(25,-5);(25,-30)**\crv{~*=<2pt>{.}(20,-17.5)}?(.6)*\dir{>}
,(30,-25)*\cir(0.3,0){}*\frm{*}
,(25,-30)*\cir(0.3,0){}*\frm{*}
,(25,-30);(30,-25)**\dir{--}?(.4)*\dir{<}
} \\ \\
Figure 3: Krajewski diagram of the standard model with right-handed  neutrino \\
\hskip-1.1cm and Majorana-mass term depicted by the dashed arrow.
\\
\end{tabular}
\end{center}

Certainly the most restrictive constraint on the discrete parameters  concerns the scalar representation. In the Yang-Mills-Higgs ansatz it  is an arbitrary input. In the almost commutative setting it is {\it  computed} from the data of the inner spectral triple.

We find it hard to believe in a coincidence if, despite the mentioned  constraints, the standard model fits perfectly into the almost  commutative frame. On the other hand, no left-right symmetric model  does \cite{lr}, no grand unified theory does \cite{gut} and we have  no supersymmetric model that does \cite{ss}, see figure 4.
\addtocounter{figure}{1}
\begin{figure}[h]
\label{versus}
\epsfxsize=11cm
\hspace{2.1cm}
\epsfbox{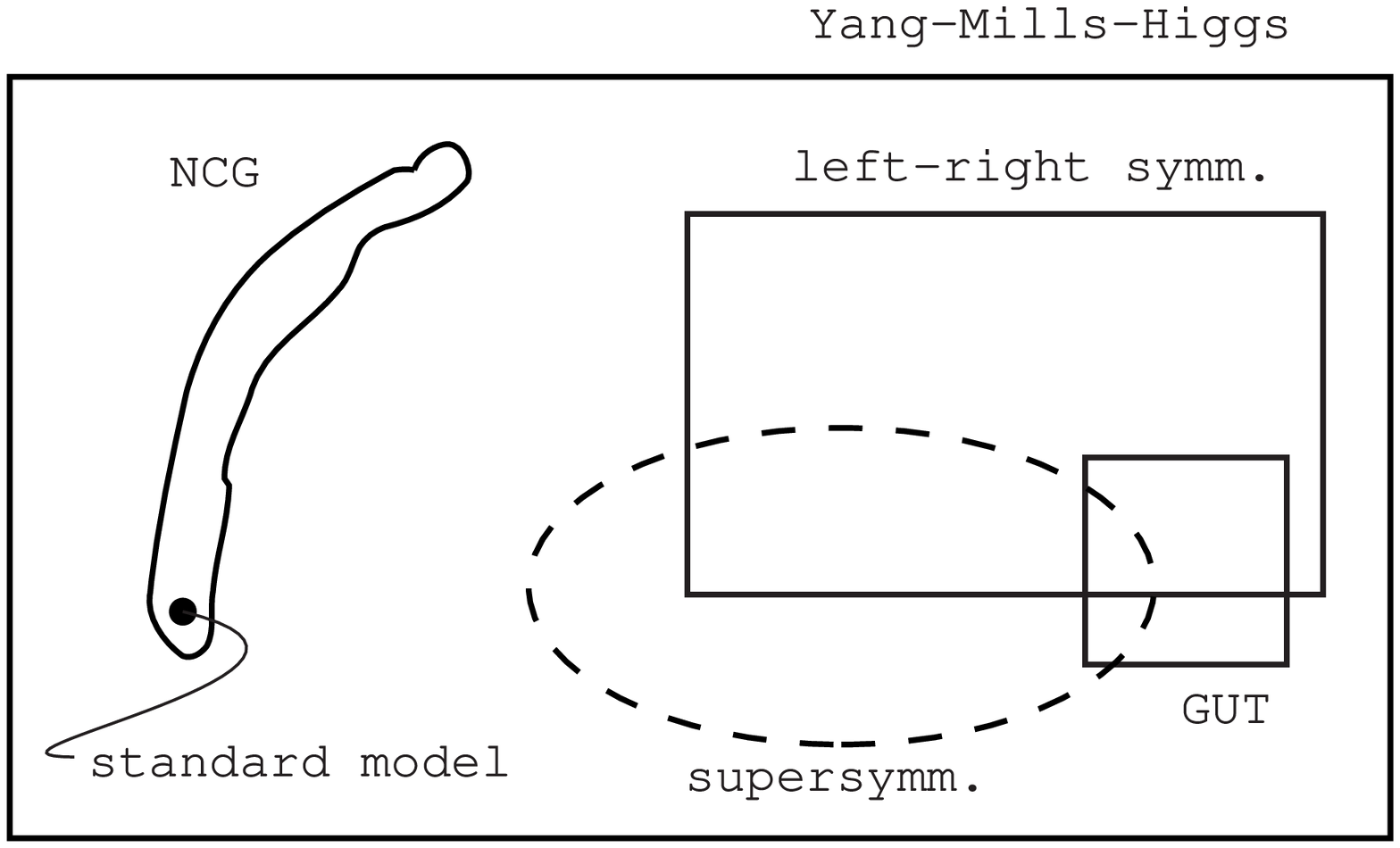}
\caption{The almost commutative model building kit}
\end{figure}

Here is the inner triple of the standard model with one generation of  fermions including a right-handed neutrino. The algebra has four  summands:
\bb\aaa=\hhh\op\cc\op M_3(\cc)\op\cc\owns (a,b,c,d),\ee
the Hilbert space is 32-dimensional
\bb
\hh=\hh_L\op\hh_R\op\hh_L^c\op\hh_R^c,\ee
\bb
\hh_L=\hh_L^c\!\!&\!\!=\!\!&\!\!(\cc^2\ot \cc^3)\,\op\,
(\cc^2\ot \cc),\\
\hh_R=\hh_R^c\!\!&\!\!=\!\!&\!\!(\cc\ot \cc^3)\,\op\,
(\cc\ot \cc^3)\,\op\,
(\cc\ot \cc)\,\op\,(\cc\ot \cc) ,\ee
and carries the faithful repesentation
\bb \rho(a,b,c):=
\pp{\rho_{L}&0&0&0\cr
0&\rho_{R}&0&0\cr
0&0&{\bar\rho^c_{L}}&0\cr
0&0&0&{\bar\rho^c_{R}}}\ee
with
\bb\rho_{L}(a):=\pp{
a\ot 1_3&0\cr
0&a},\
\rho_{R}(b):= \pp{
b  1_3&0&0&0\cr 0&\bar b  1_3&0&0\cr
0&0&b&0\cr
0&0&0& d}, \label{repr1}
\ee\bb
\rho^c_{L}(c,d):=\pp{
1_2\ot c&0\cr
0& \bar d1_2},\
\rho^c_{R}(c,d) := \pp{
c&0&0&0\cr 0& c&0&0\cr
0&0&\bar d&0\cr
0&0&0&\bar d}.  \label{repr2}
\ee
The Dirac operator
\bb \tilde \dd=\pp{0&\mm&0&0\cr
\mm^*&0&0&S\cr
0&0&0&\bar\mm\cr
0&S^*&\bar\mm^*&0}\ee
contains Dirac masses
\bb\mm=\pp{
\pp{M_u&0\cr 0&0} \ot 1_3+
\pp{0&0\cr 0&M_d}\ot 1_3
&0\cr
0&\pp{M_\nu&0\cr 0&0}+
\pp{0&0\cr 0&M_e}}\ee
and the Majorana mass for the right-handed neutrino,
\bb S=\pp{M_M&0\cr0&0} .\ee
For $N$ generations $M_u,\ M_d,\ M_\nu$ and $M_e$ are complex $N \times N$ matrices encoding the Dirac masses and mixings while
$M_M$ is a complex, symmetric matrix encoding Majorana masses and  mixings. For example for the quarks with $N=3$ generations:
\bb M_u:=\pp{
m_u&0&0\cr
0&m_c&0\cr
0&0&m_t},&& M_d:= C_{KM}\pp{
m_d&0&0\cr
0&m_s&0\cr
0&0&m_b},\ee
with the Cabibbo-Kobayashi-Maskawa matrix  $C_{KM}$.

This model is conform with the standard formulation of the axiom of Poincar\'e
duality as stated in \cite{book}. It seems to be closely related to the
older bi-module approach of the Connes-Lott model \cite{lott} which
also exhibits two copies of the complex numbers in the algebra.
In the original version of the standard model with right-handed Majorana
neutrinos Chamseddine, Connes \& Marcoli \cite{mc2} used an
alternative spectral triple based on the algebra $\aaa=\hhh\op\cc\op M_3(\cc)$.
But this spectral triple requires a
subtle change in the formulation of the Poincar\'e duality, i.e. it needs
two elements to generate $KO$-homology as a module over $K_0$.
It should however be noted that right-handed neutrinos that allow for
Majorana-masses always fail to fulfil the axiom of orientability \cite{ko6}
since the representation of the algebra does not allow to construct
a Hochschild-cycle reproducing the chirality operator. For this reason we have drawn the arrows connected to the right-handed neutrinos with broken lines in the Krajewski diagram.

\subsection{Constraints on the continuous parameters}

\subsubsection{The dimensionless parameters}

The spectral action counts the number of eigenvalues of the Dirac  operator whose absolute values are less than the energy cut-off $ \Lambda$. On the input side its continuous parameters are: this cut- off, three positive parameters in the cut-off functions and the  parameters of the inner Dirac operator, i.e. fermion masses and  mixing angles. On the output side we have: the cosmological constant,  Newton's constant, the gauge and the Higgs couplings. Therefore there  are constraints, which for the standard model with $N$ generations  and three colours read:
\bb \,\frac{5}{3}\,g_1^2=  g_2^2=g_3^2=\,\frac{3}{N}\,\frac{Y_2^2}{H} \,\frac{\lambda}{24}\,= \,\frac{3}{4N}\,Y_2\,.\label{4con}\ee
Here $Y_2$ is the sum of all Yukawa couplings $g_f$ squared, $H$ is  the sum of all Yukawa couplings raised to the fourth power. Our  normalisations are: $m_f=\sqrt{2}\,(g_f/g_2)\,m_W,$ $(1/2)\,(\pa  \varphi)^2+(\lambda/24)\,\varphi^4$. If we define the gauge group by  lifted automorphisms rather than unimodular unitairies, then we get  an ambiguity parameterized by the central charges. This ambiguity  leaves the $U(1)$ coupling $g_1$ unconstrained and therefore kills  the first of the four constraints (\ref{4con}).

Note that the noncommutative constraints (\ref{4con}) are different
from Veltman's condition \cite{velt}, which in our normalisation reads:

\bb  {\textstyle\frac{3}{4}}g_2^2+{\textstyle\frac{1}{4}}g_1^2 + {\textstyle\frac{1}{3}}\lambda-2g_t^2=0.\ee

Of course the constraints (\ref{4con}) are not stable under the  renormalisation group flow and as in grand unified theories we can  only interpret them at an extremely high unification energy $\Lambda $. But in order to compute the evolution of the couplings between our  energies and $\Lambda$ we must resort to the daring hypothesis of the  big desert, popular since grand unification. It says: above presently  explored energies and up to $\Lambda$ no more new particle, no more  new forces exist, with the exception of the Higgs, and that all  couplings evolve without leaving the perturbative domain. In  particular the Higgs self-coupling $\lambda$ must remain positif. In  grand unified theories one believes that new particles exist with  masses of the order of $\Lambda$, the leptoquarks. They mediate  proton decay and stabilize the constraints between the gauge  couplings by a bigger group. In the noncommutative approach we  believe that at the energy $\Lambda$ the noncommutative character of  space-time ceases to be negligible. The ensuing uncertainty relation  in space-time might cure the short distance divergencies and thereby  stabilize the constraints. Indeed Grosse \& Wulkenhaar have an  example of a scalar field theory on a noncommutative space-time with  vanishing $\beta$-function \cite{raimar}.

Let us now use the one-loop $\beta$-functions of the standard model   with $N=3$ generations to evolve the constraints (\ref{4con}) from $E= \Lambda$  down to our energies $E=m_Z$. We set:
$ t:=\ln (E/m_Z),\qq \de g/\de t=:\beta _g,\qq \kappa :=(4\pi )^{-2}. $ We will neglect all fermion masses below the top mass and also  neglect threshold effects. We admit a Dirac mass $m_D$ for the $\tau$  neutrino induced by spontaneous symmetry breaking and take this mass  of the order of the top mass. We also admit a Majorana mass $m_M$ for  the right-handed $\tau$ neutrino. Since this mass is gauge invariant  it is natural to take it of the order of $\Lambda$. Then we get two  physical masses for the
$\tau$ neutrino: one is tiny, $m_\ell\sim m_D^2/m_M$, the other is  huge, $m_r\sim m_M.$ This is the popular seesaw mechanism \cite {seesaw}. The renormalisation of these masses is well-known \cite {seesawren}. By the Appelquist-Carazzone decoupling theorem \cite{ac}  we distinguish two energy domains: $E>m_M$ and $E<m_M$. In the  latter, the Yukawa coupling of the $\tau$ neutrino drops out of the $ \beta$-functions and is replaced by an effective coupling
\bb k=2\,\frac{g_\nu^2}{m_M}\,, \qq{\rm at\ }E=m_M.\ee
At high energies, $E>m_M$, the $\beta$-functions are \cite{mv,jones}:
\bb \beta _{g_i}&=&\kappa b_ig_i^3,\qq b_i=
{\textstyle
\left( \frac{20}{9} N+\frac{1}{6},-\frac{22}{3}+\frac{4}{3} N+\frac{1} {6},
-11+\frac{4}{3} N\right) },
\\ \cr
\beta _t&=&\kappa
\left[ -\sum_i c_i^ug_i^2 +Y_2 +\,\frac{3}{2}\,g_t^2
\,\right] g_t,\\
\beta _\nu&=&\kappa
\left[ -\sum_i c_i^\nu g_i^2 +Y_2 +\,\frac{3}{2}\,g_\nu^2\,
\right] g_\nu,\\
\beta _\lambda &=&\kappa
\left[ \,\frac{9}{4}\,\left( g_1^4+2g_1^2g_2^2+3g_2^4\right)
-\left( 3g_1^2+9g_2^2\right) \lambda
+4Y_2\lambda -12H+4\lambda ^2\right] ,\ee
with
\bb c_i^t=\left( {\textstyle\frac{17}{12}},{\textstyle\frac{9}{4}} , 8 \right) ,
&
c_i^\nu =\left( {\textstyle\frac{3}{4}},{\textstyle\frac{9}{4}} , 0 \right) ,\\
Y_2=3g_t^2+g_\nu^2,
&
H=3g_t^4+g_\nu^4.
\ee
At low energies, $E<m_M$,
the $\beta$-functions are the same except that
$Y_2=3g_t^2$, $H=3g_t^4$ and that $\beta_\nu$ is replaced \cite {seesawren} by:
\bb
\beta _k=\kappa
\left[ -3g_2^2+\,\frac{3}{2}\,-\sum_i c_i^\nu g_i^2 +Y_2 +\,\frac{2} {3}\,\lambda+2Y_2\,
\right] k.\ee
We suppose that all couplings (other than $g_\nu$ and $k$) are  continuous  at $E=m_M$, no threshold effects.
The three gauge couplings decouple from the other equations and have  identical evolutions in both energy domains:
\bb g_i(t)=g_{i0}/\sqrt{1-2\kappa b_ig_{i0}^2t}.\ee
The initial conditions are taken from experiment \cite{data}:
\bb g_{10}= 0.3575,\qq
g_{20}=0.6514,\qq
g_{30}=1.221.\ee
In a first run we leave $g_1$ unconstrained. Then the unification  scale $\Lambda $ is the solution of $g_2(\ln (\Lambda /m_Z))=g_3(\ln  (\Lambda /m_Z))$,
\bb \Lambda = m_Z\exp\frac{g_{20}^{-2}-g_{30}^{-2}}{2\kappa (b_2-b_3)} \,=\,1.1\times 10^{17}\  {\rm GeV},\ee
and is independent of the number of generations.

Then we choose $g_\nu=Rg_t$ at $E=\Lambda$ and $m_M$ and solve  numerically the evolution equations for $\lambda,\ g_t,\ g_\nu$ and $k $ with initial conditions at $E=\Lambda$ from the noncommutative  constraints (\ref{4con}):
\bb g_2^2=\,\frac{3+R^2}{3+R^4}\,\frac{\lambda}{24}\,= \,\frac{3+R^2} {4}\,g_t^2\,.\ee
We note that these constraints imply that all couplings remain  perturbative and
at our energies we obtain the pole masses of the Higgs, the top and  the light neutrino:
\bb m_H^2=\,\frac{4}{3}\,\frac{\lambda(m_H) }{g_2(m_Z)^2}\,m_W^2,\qq
m_t=\sqrt{2}\,\frac{g_t(m_t)}{g_2(m_t)}\,m_W,\qq
m_\ell=\,\frac{k(m_Z)}{g_2(m_Z)^2}\,m_W^2.\ee
A few numerical results are collected in table 1.
\begin{table}[h]
\begin{center}
\begin{tabular}{|c|c|c|c|c|c|c|c|c|c|}
\hline
$g_\nu/g_t|_\Lambda$ & 0 & 1.16& 1.16&1.2&1.2&1.3&1.3&1.4&1.4
\\
$m_M$ [GeV]& *&$2\cdot10^{13}$&$10^{14}$&$2\cdot10^{13}$&$10^{14}$&$3 \cdot10^{13}$&$10^{14}$&$3\cdot10^{13}$&$10^{14}$\\
\hline
$m_t$ [GeV]&186.3&173.3&173.6&172.5&172.8&170.5&170.7&168.4&168.6\\
$m_H$ [GeV]&188.4&170.5&170.8&169.7&170.0&167.7&168.0&165.8&166.1\\
$m_\ell$ [ eV]&0&0.29&0.06&0.30&0.06&0.23&0.07&0.25&0.08\\
\hline
\end{tabular}
\end{center}
\caption{The top, Higgs and neutrino masses as a function of $R$ and  of the Majorana mass for the unification scale $\Lambda = 1.1\times  10^{17}$ GeV}
\end{table}

Note that the Higgs mass is not very sensitive to the three input  parameters,  $\Lambda,\ m_M,$ and $ R=g_\nu(\Lambda)/g_t(\Lambda)$ as  long as they vary in a range reproducing senible masses for the top  and the light neutrino, today $m_t=170.9\,\pm2.6$ GeV and $0.05\ {\rm  eV}\ <\ m_\ell\ <\ 0.3$ eV. Then we have for the Higgs mass
\bb m_H= 168.3\pm 2.5\ {\rm GeV}.\ee

In a second run we use the constraint on the Abelian coupling $\,\frac {5}{3}\,g_1^2=  g_2^2$
to compute the unification scale: \bb  \Lambda = m_Z\exp\frac{g_{20}^ {-2}-(3/5)g_{30}^{-2}}{2\kappa (b_2-(3/5)b_1)}\,=\,9.8\times 10^{12} \  {\rm GeV}.\ee

Note that the third constraint $\,\frac{5}{3}\,g_1^2=  g_3^2$ yields  an intermediate unification scale, $\Lambda =2.4\times 10^{14}\  {\rm  GeV}.$ Again we give a few numerical results, table 2.
\begin{table}[h]
\begin{center}
\begin{tabular}{|c|c|c|c|c|c|}
\hline
$g_\nu/g_t|_\Lambda$ & 0 & 0.95& 1&1.1&1.2
\\
\hline
$m_t$ [GeV]&183.4&173.3&172.3&170.3&168.1\\
$m_H$ [GeV]&188.5&174.9&173.8&171.9&170.2\\
$m_\ell$ [ eV]&0&0.53&0.57&0.67&0.77\\
\hline
\end{tabular}
\end{center}
\caption{The top, Higgs and neutrino masses as a function of $R$ for  the Majorana mass and the unification scale $m_M=\Lambda = 9.8\times  10^{12}$ GeV}
\end{table}

Here the upper bound for the light neutrino mass cannot be met  strictly with $m_M<\Lambda$. This does not worry us because that  bound derives from cosmological hypotheses.
Honouring the constraints for all three gauge couplings then yields  the combined range for the Higgs mass,
\bb m_H= 168.3^{+6.6}_{-2.5}\ {\rm GeV}.\ee

\subsubsection{Constraints on the dimensionful parameters}
The spectral action also produces constraints between the quadratic  Higgs coupling, the Planck mass, $m_P^2=1/G$,
and the cosmological constant in terms of the cut-off $\Lambda$ and  of the three moments, $f_0, f_2, f_4$, of the cut-off function. A  step function for example has $2f_0=f_2= f_4$.  Trading the quadratic  Higgs coupling for the $W$ mass, these constraints read:
\bb m_W^2=\frac{45}{4\cdot96}\frac{(3m_t^2+m_\nu^2)^2}{3m_t^4+m_\nu^4} \left[\frac{f_2}{f_4}\,\Lambda^2-
\frac{m_\nu^2}{3m_t^2+m_\nu^2}\,m_M^2\right]
\sim \frac{45}{96}\left[\frac{f_2}{f_4}\,\Lambda^2-
\frac{1}{4}\,m_M^2\right],
\ee\bb
m_P^2&=&\frac{1}{3\pi}\left[
\left(96-4\,\frac{(3m_t^2+m_\nu^2)^2}{3m_t^4+m_\nu^4}\right)f_2\Lambda^2
+2\left(2\,\frac{m_\nu^2(3m_t^2+m_\nu^2)}{3m_t^4+m_\nu^4}-1\right) f_4m_M^2\right]\cr\cr
&\sim&
\frac{1}{3\pi}\left[80 f_2\Lambda^2
+2 f_4m_M^2\right],
\\[1mm]
\Lambda_c&\sim&\frac{1}{\pi m_P^2}[(96\cdot 2 f_0-16 f_22/f_4) \Lambda^4+f_4m_M^4+4f_2\Lambda^2m_M^2].\ee
We have taken three generations, i.e. a 96-dimensional inner Hilbert  space, we only kept the Yukawa couplings of the  top quark and of the  $\tau$ neutrino, and one single Majorana mass in the third  generation. The experts still do not agree whether the  renormalisation  group flow of the quadratic Higgs coupling is  logarithmic or quadratic in the energy $E$. Nobody knows how Newton's  and the cosmological constants depend on energy. Therefore we cannot  put the above constraints to test. It is however reassuring that,  thanks to the seesaw mechanism, a $W$ mass much smaller than the  Planck mass is easily obtained. On the other hand it is difficult to  produce a small cosmological constant.

\section{Is the standard model special?}
\subsection{Privileged solutions of the constraints on the discrete  parameters}

Despite all constraints, there is still an infinite number of Yang- Mills-Higgs-Dirac-Yukawa models that can be derived from gravity  using almost commutative geometry. The exploration of this special  class is highly non-trivial and starts with Krajewski diagrams. Of  course one would like to show that the standard model has a  privileged position in the class as indicated in figure 4.

At present there are two approaches in this direction.

The first by Chamseddine, Connes \& Marcolli \cite{mc2} starts from a  left-right symmetric algebra. This algebra admits a privileged bi-module which is identical to the fermionic Hilbert space of the  standard model. The algebra of the standard model is a maximal  subalgebra of the left-right symmetric one and the inner Dirac  operator is almost the maximal operator satisfying the axioms of a  spectral triple. The number of colours and the number of generations  remain unexplained in this approach.

The second approach again has nothing to say about the number of  colours and generations. It is a more opportunistic approach and  copies what grand unified theories \cite{gglash} did in the frame of  Yang-Mills-Higgs-Dirac-Yukawa theories. There, the idea was to cut  down on the number of possible
models with a `shopping' list of requirements: one wants
\begin{itemize}\item
the gauge group to be simple,
\item
the fermion representation irreducible,
\item
the fermion representation complex under the gauge group,
\item
the fermion representation to be free of Yang-Mills anomalies,
\item
the model to contain the standard model.
\end{itemize}
The motivations for these requirements were of heteroclitic origin,
mathematical simplicity, the wish to be able to distinguish particle
from anti-particles, consistency of the ensuing quantum field theory,
consistency with the phenomenology. Two examples with gauge group $SO (10)$ and $E_6$ and accounting for one generation of fermions were on  the market. On the other hand at that time $F_4$ was banned because  it does not have complex representations.

Coming back to Connes' noncommutative model building kit, we remark  that the spectral triple of the standard model with one generation of  fermions and a massless neutrino is irreducible. It has another  remarkable property concerning its built-in spontaneous symmetry  breaking: it allows a vacuum giving different masses to the two  quarks although they sit in an isospin doublet. Indeed, in the  majority of noncommutative models the spontaneous symmetry breaking  gives degenerate masses to fermions in irreducible {\it group}  representations. We say that those models are `` dynamically  degenerate'' because there are other mass degeneracies coming from  the kinematics, i.e the axioms of the spectral triple without using  the spectral action and its induced spontaneous symmetry breaking.  These ``kinematical'' degeneracies are always protected by a gauge  group, which we call ``colour group'' because in the standard model  this group is the colour group.

For years we have been looking for viable noncommutative models other  than the standard model, without success. We therefore started to  scan the Krajewski diagrams with the following shopping list. We want:
\begin{itemize}\item
the spectral triple to be irreducible,
\item
the fermion representation to be complex under the little group in  every irreducible component,
\item
possible massless fermions to transform trivially under the little  group,
\item
the fermion representation to be free of Yang-Mills and mixed  gravitational Yang-Mills anomalies,
\item
the spectral triple to have no dynamical degeneracy and the colour  group of every kinematical degeneracy to remain unbroken.
\end{itemize}

The first step is to get the list of irreducible Krajewski diagrams  (letter-changing arrows only) \cite{class}. In the case of an inner  spectral triple of Euclidean signature, we have no such diagram for a  simple algebra, one diagram for an algebra with two simple summands,  30 diagrams for three summands, 22 diagrams for four summands,  altogether 53 irreducible diagrams for algebras with up to four  simple summands. The situation simplifies when we go to the  Lorentzian signature where we remain with only 7 diagrams for up to  four summands. These numbers are summarized in table 3.

\begin{table}[h]
\begin{center}
\begin{tabular}{|c|c|c|}
\hline\\[-3mm]
\#(summands) & Eulidean & Lorentzian
\\[1ex]
\hline
1&$\ \ 0$&0\\
2&$\ \ 1$&0\\
3& 30&0\\
4& 22&7\\
$\!\!\!\!\!\!\le 4$& 53&7\\
\hline
\end{tabular}
\end{center}
\caption{The number of irreducible Krajewski diagrams for algebras  with up to 4 simple summands and inner spaces with Eucidean and  Lorentzian signatures, letter-changing arrows only }
\end{table}

The second step is to scan all models derived from the irreducible  Krajewski diagrams with respect to our shopping list. In both  signatures we remain with the following models:
The standard model with one generation of fermions, an arbitrary  number of colours $C\ge 2$ and a massless neutrino:
\bb\frac{SU(2)\times U(1)\times SU(C)}{\zz_2\times \zz_C}&
\longrightarrow
&\frac{U(1)\times SU(C)}{\zz_C}\ee
For even $C$ the $\zz_2$ factor is missing. We also have three  possible submodels with identical fermion content, but with $SU(2)$  replaced by $SO(2)$, no $W$-bosons, or with $SU(C)$ replaced by $SO(C) $ or $USp(C/2)$, $C$ even, less gluons. There is one more possible  model, the elctro-strong model:
\bb{ U(1)\times SU(C)}&
\longrightarrow
&{U(1)\times SU(C)}\ee
The fermionic content is ${\ul C}\op {\ul 1}$, one quark and
one charged lepton.  The two electric charges
are arbitrary but vectorlike. The model has no scalar and no symmetry  breaking.

For those of you who think that our shopping list is unreasonably  restrictive already in the frame of Yang-Mills-Higgs-Dirac-Yukawa  models, here is a large class of such models satisfying our shopping  list: Take any group that has complex representations (like $E_6$)  and take any irreducible, complex, unitary representation of this  group. Put the left- and right-handed fermions in two copies of this  representation, choose the Hilbert space for scalars 0-dimensional  and a gauge-invariant mass for all fermions.

\subsection{Beyond the standard model}
For many years we have been trying to construct models from  noncommutative geometry that go beyond the standard model \cite {beyond} and we failed to come up with anything physical if it was  not to add more generations and right-handed neutrinos to the  standard model.

The noncommutative constraints on the continuous parameters of the  standard model with $N=4$ generations fail to be compatible with the  hypothesis of the big desert \cite{knecht}.

Since a computer program \cite{prog} was written to list the  irreducible Krajewski diagrams for algebras with more than three  summands we do have a genuine extension of the standard model  satisfying all physical desirata. It comes from an algebra with six  summands \cite{chris} and is identical to the standard model with two  additional leptons $A^{--}$ and $C^{++}$ whose electric charge is two  in units of the electron charge. These new leptons  couple neither to  the charged gauge bosons, nor to the Higgs scalar. Their hypercharges  are vector-like, so that they do not contribute to the electroweak  gauge anomalies. Their masses are gauge-invariant and they constitute  viable candidates for cold dark matter \cite{klop}.

Also, by trial and error, two more models could be found recently \cite{colour,vector}.
The first model is based on an algebra with six summands and adds to the
standard model a lepton-like, weakly charged, left-handed doublet and two right-handed
hypercharge singlets. These four particles are each colour-doublets under
a new $SU(2)_c$ colour group. They participate in the
Higgs mechanism and the noncommutative constraints require masses
to be around $75$ GeV. Since they have a non-Abelian colour group one
expects a macroscopic confinement \cite{okun} with a confinement radius
of $\sim 10^{-5}$ cm. Although these particles have electro-magnetic
charge after symmetry breaking it is not yet clear whether they could have
been detected in existing experiments.

The second model adds to the
standard model three generations of vectorlike doublets with weak and
hypercharge. After symmetry breaking one particle of each doublet becomes
electrically neutral while its partner acquires a positive or a negative
electro-magnetic charge (depending on the choice of the hypercharge).
The particles, like the $AC$ leptons, do not couple to the Higgs boson
and should have masses of the order of $\sim 10^3$ TeV.  Due to
differing self-interaction terms with the photon and the Z-boson the
neutral particle will be slightly lighter than the charged particle and
is therefore the stable state \cite{wells}. Together with its neutrino-like cross section
the neutral particle constitutes an interesting dark matter candidate \cite{griest}.

Another interesting aspect of the models presented in \cite{colour}
and \cite{vector} is that they naturally exhibit gauge unification
at around $10^{13}$ GeV. This should be seen in contrast to the $AC$ model
which only aggravates the situation of the lacking unification in
the standard model.

\section{Conclusions}

There are two clear-cut achievements of noncommutative geometry in  particle physics:
\begin{itemize}\item
Connes' derivation of the Yang-Mills-Higgs-Dirac-Yukawa ansatz from  the Einstein-Hilbert action,
\item
the fact that this unification of all fundamental forces allows to  compute correctly the representation content of the Higgs scalar  (i.e. one weak isospin doublet, colour singlet with hyper-charge one  half) from the experimentally measured representation content of the  fermions.
\end{itemize}
The other clear achievements are restrictions on the gauge groups,  severe restrictions on the fermion representations and their  compatibility with experiment.

Finally there are constraints on the top and Higgs masses. They do  rely on the hypothesis of the big desert. Nevertheless we look  forward to the Tevatron and LHC verdict.

To our taste the comparison of these achievements with the  explanation of the Balmer-Rydberg formula
by quantum mechanics is fair. We think of the early Bohr model, which  already did this job. The Bohr model certainly had its shortcomings  and did not pretend to be the last word, but it pointed in the right  direction. So in what direction is noncommutative geometry pointing?  For us it holds the promise to clean up quantum field theory and  thereby including quantum gravity. Apparently this motivation for  noncommutative geometry already goes back to Heisenberg \cite{jackiw}.

\end{document}